\documentclass[11pt]{article}
\newcommand{\ket}[1]{\left | #1 \right \rangle}
\newcommand{\bra}[1]{\left \langle #1 \right |}

\def\openone{\leavevmode\hbox{\small1\kern-3.8pt\normalsize1}}
\def\RR{{\rm I\kern-.2emR}}

\newcommand{\proj}[1]{\ket{#1}\!\bra{#1}}

\newcommand{\inner}[2]{ \langle #1 | #2 \rangle}

\newcommand{\beq}{\begin{equation}}
\newcommand{\eeq}{\end{equation}}
\newcommand{\beqa}{\begin{eqnarray}}
\newcommand{\eeqa}{\end{eqnarray}}

\newtheorem{theorem}{Theorem}

\newtheorem{lemma}{Lemma}

\begin{document}
\begin{center}
{\LARGE\bf A stronger no-cloning theorem\\ }
\bigskip
{\normalsize Richard Jozsa}\\
{\small\it Department of Computer Science, University of
Bristol,\\ Merchant Venturers Building, Bristol BS8 1UB U.K.}
\\[4mm]
\date{today}
\end{center}

\begin{abstract}
It is well known that (non-orthogonal) pure states cannot be
cloned so one may ask: how much or what kind of additional
(quantum) information is needed to supplement one copy of a
quantum state in order to be able to produce two copies of that
state by a physical operation? For classical information, no
supplementary information is required. However for pure quantum
(non-orthogonal) states, we show that the supplementary
information must always be as large as it can possibly be i.e. the
clone must be able to be generated from the additional information
alone, independently of the first (given) copy.

\end{abstract}
\section{Introduction}
It is well known that (non-orthogonal) pure states cannot be
cloned \cite{WZ} i.e. if $\{ \ket{\psi_i}\}$ is a set of pure
states containing at least one non-orthogonal pair then no
physical operation can achieve the transformation $ \ket{\psi_i}
\longrightarrow \ket{\psi_i}\ket{\psi_i}$. Evidently two copies of
a quantum state contain strictly more ``information'' about the
state than is available from just one copy so in view of the
impossibility of cloning one may ask: what additional information
is needed to supplement one copy $\ket{\psi_i}$ in order to be
able to produce two copies $\ket{\psi_i}\ket{\psi_i}$? For
classical information, no supplementary information at all is
needed and one might guess that as the set $\{ \ket{\psi_i}\}$
becomes ``more classical'' the necessary supplementary information
should decrease in some suitable way. However we prove below that
this is not the case: we show (for mutually non-orthogonal states)
that the supplementary information must always be as large as it
can possibly be i.e. the second copy $\ket{\psi_i}$ can always be
generated from the supplementary information alone, independently
of the first (given) copy. Thus in effect, cloning of
$\ket{\psi_i}$ is possible only if the second copy is fully
provided as an additional input.

Using our techniques we will also give simple proof of the
Pati-Braunstein no-deleting principle \cite{PB} (for the mutually
non-orthogonal states) and discuss its relation to cloning.
\section{No cloning}
We now give a precise formulation of our main result. By a
physical operation we will mean a trace preserving completely
positive map. Note that this definition excludes the collapse of
wavefunction in a quantum measurement, as a valid physical
process. (This will be relevant to our later discussion of the
no-deleting principle.) {\theorem \label{thm1} Let $\{
\ket{\psi_i}\}$ be any finite set of pure states containing no
orthogonal pairs of states. Let $\{ \rho_i \}$ be any other set of
(generally mixed) states indexed by the same labels. Then there is
a physical operation \[ \ket{\psi_i}\otimes \rho_i \longrightarrow
\ket{\psi_i}\ket{\psi_i} \] if and only if there is a physical
operation \[ \rho_i \longrightarrow \ket{\psi_i} \] i.e. the full
information of the clone must already be provided in the ancilla
state $\rho_i$ alone.}

\noindent {\bf Remark}\cite{errors}. If the set $\{ \ket{\psi_i}
\}$ contains some orthogonal pairs then the unassisted clonability
of orthogonal states spoils the simplicity of the statement of
theorem 1. As an example consider
\[ \begin{array}{ll} \ket{\psi_1}=\ket{0} &
\ket{\alpha_1}=\ket{a}  \\   \ket{\psi_2}=\ket{1} &
\ket{\alpha_2}=\ket{a} \\ \ket{\psi_3} = \frac{1}{\sqrt{2}}
(\ket{0}+\ket{1}) & \ket{\alpha_3}=\ket{b}
\end{array}
\]
where $\ket{a}$ and $\ket{b}$ are orthogonal. Then clearly
$\ket{\psi_i}\ket{\alpha_i} \rightarrow \ket{\psi_i}\ket{\psi_i}$
is possible (as $\{ \ket{\psi_i}\ket{\alpha_i} \}$ is an
orthonormal set) but $\ket{\alpha_i}\rightarrow \ket{\psi_i}$ is
not possible (as $\ket{\alpha_1}=\ket{\alpha_2}$ but
$\ket{\psi_1}\neq \ket{\psi_2}$). Indeed the $\ket{\alpha_i}$
states here provide reliable distinguishability of $i$ values in
the case that this is not already provided by the $\ket{\psi_i}$'s
themselves. $QED$.

To prove the theorem we will use the following lemma which is
proved as lemma 1 of \cite{JS}. {\lemma \label{lemma1} Let $\{
\ket{\alpha_i} \}$ and $\{ \ket{\beta_i} \}$ be two sets of pure
states (indexed by the same labels). Then the two sets have equal
matrices of inner products (i.e.
$\inner{\alpha_i}{\alpha_j}=\inner{\beta_i}{\beta_j}$ for all $i$
and $j$) if and only if the two sets are unitarily equivalent
(i.e. there exists a unitary operation $U$ on the direct sum of
the state spaces of the two sets with
$U\ket{\alpha_i}=\ket{\beta_i}$ for all $i$). }

\noindent {\bf Proof of theorem \ref{thm1}} \,\, Suppose that
there is a physical operation $\rho_i \rightarrow \ket{\psi_i}$.
Then clearly $\ket{\psi_i}\otimes \rho_i \rightarrow
\ket{\psi_i}\ket{\psi_i}$ is allowed.

Conversely suppose that there is a physical operation
\begin{equation} \label{cl}
\ket{\psi_i}\otimes \rho_i \longrightarrow
\ket{\psi_i}\ket{\psi_i}. \end{equation} Consider first the case
that $\rho_i$ are pure states, $\ket{\alpha_i}$ say. The physical
operation eq. (\ref{cl}) may be expressed as a unitary operation
if we include an environment space, initially in a fixed state
denoted $\ket{A}$. For clarity we include an extra register,
initially in a fixed state $\ket{0}$, that is to receive the clone
of $\ket{\psi_i}$. Then eq. (\ref{cl}) may be written as a {\em
unitary} transformation \[
\ket{\psi_i}\ket{0}\ket{\alpha_i}\ket{A} \longrightarrow
\ket{\psi_i}\ket{\psi_i}\ket{C_i} \] where $\ket{C_i}$ (generally
depending on $i$) is the output state of the two registers that
initially contained $\ket{\alpha_i}\ket{A}$. Hence by unitarity,
the two sets $\{ \ket{\psi_i}\ket{\alpha_i} \}$ and $\{
\ket{\psi_i}\ket{\psi_i}\ket{C_i} \}$ have equal matrices of inner
products and then, so do the sets $\{ \ket{\alpha_i} \} $ and $\{
\ket{\psi_i} \ket{C_i} \}$ (by a simple cancellation of
$\inner{\psi_i}{\psi_j}$ from the two initial matrices). Thus by
lemma \ref{lemma1} these two sets are unitarily equivalent so
$\ket{\psi_i}$ can be generated from $\ket{\alpha_i}$ alone (by
applying the unitary equivalence and discarding the $\ket{C_i}$
register).

If $\rho_i$ are mixed we express them as probabilistic mixtures of
pure states \[ \rho_i = \sum_k p^{(i)}_k \proj{\alpha^{(i)}_k}. \]
Then a physical operation achieves \[ \ket{\psi_i}\otimes \rho_i
\longrightarrow \ket{\psi_i} \ket{\psi_i} \hspace{1cm} \mbox{for
all $i$}
\] iff it achieves \begin{equation} \label{eq1}
 \ket{\psi_i}\otimes \ket{\alpha^{(i)}_k}
\longrightarrow \ket{\psi_i} \ket{\psi_i} \hspace{1cm} \mbox{for
all $i$ and $k$.} \end{equation}  By the pure state analysis
above, a physical operation effecting eq. (\ref{eq1}) exists iff
there is a physical operation effecting \[ \ket{\alpha^{(i)}_k}
\longrightarrow \ket{\psi_i} \hspace{1cm} \mbox{for all $i$ and
$k$} \] and then we get $\rho_i \longrightarrow \ket{\psi_i}$ too.
$QED$.

In the particular case of cloning assisted by {\em classical}
information (i.e. the states $\rho_i$ are required to be mutually
commuting) we deduce that this supplementary data must contain the
full identity of the states as classical information (rather than
just quantum information). Indeed if the $\rho_i$ are classical
then they can be copied any number of times so if we can make one
clone of $\ket{\psi_i}$ from $\rho_i$, we can make arbitrarily
many clones and hence determine the the identity of $\ket{\psi_i}$
i.e. the classical information of the label $i$ must be contained
in the supplementary classical information.
\section{No deleting}
Our techniques may also be used to give a simple proof of the
Pati-Braunstein no-deleting principle \cite{PB} for sets $\{
\ket{\psi_i} \}$ that contain no orthogonal pairs. The issue here
is the following. Suppose we have two copies
$\ket{\psi_i}\ket{\psi_i}$ of a state and we wish to delete one
copy by a physical operation:
\begin{equation} \label{nd1} \ket{\psi_i}\ket{\psi_i} \longrightarrow
\ket{\psi_i}\ket{0} \end{equation} (where $\ket{0}$ is any fixed
state of the second register). As before, any such physical
operation may be expressed as a unitary operation if we include an
environment space, initially in a fixed state $\ket{A}$ say. Then
eq. (\ref{nd1}) is equivalent to the unitary transformation
\begin{equation} \label{nd2} \ket{\psi_i}\ket{\psi_i}\ket{A}
\longrightarrow \ket{\psi_i}\ket{0}\ket{A_i} \end{equation} where
the final state $\ket{A_i}$ of the environment may depend on
$\ket{\psi_i}$ in general. One way of achieving this is to simply
swap (a constant) part of the environment into the second register
but then the second copy of $\ket{\psi_i}$ remains in existence
(albeit in the environment now). The no-deleting principle states
that the second copy of $\ket{\psi_i}$ can {\em never} be
``deleted'' in the sense that $\ket{\psi_i}$ can {\em always} be
resurrected from $\ket{A_i}$. Note however, that if wavefunction
collapse is also allowed as a valid physical process then deletion
is possible. (We perform a complete measurement on $\ket{\psi_i}$
and rotate the seen post-measurement state to $\ket{0}$ by a
unitary transformation that depends on the measurement outcome.)

To see the no-deleting principle with our methods, note that the
unitarity of eq. (\ref{nd2}) implies that the sets $\{
\ket{\psi_i}\ket{\psi_i}\ket{A} \}$ and $\{
\ket{\psi_i}\ket{0}\ket{A_i} \}$ have equal matrices of inner
products and then as before, so do the sets $\{ \ket{\psi_i} \}$
and $\{ \ket{A_i} \}$. Thus lemma \ref{lemma1} states that these
sets are unitarily equivalent, which is just the no-deleting
principle. \section{Further remarks} Deleting and cloning have a
common feature: in cloning we saw that the existence of the first
copy $\ket{\psi_i}$ provided no assistance in constructing the
second copy from the supplementary information. Similarly for
deletion, the existence of the first copy provides no assistance
in deleting the second copy -- in effect the only way to delete
the second copy is to transform it out into the environment (i.e.
$\ket{0}\ket{A_i} $ in eq. (\ref{nd2}) is a unitary transform of
$\ket{\psi_i}\ket{A}$ alone) and this again makes no use of the
first copy. Considering no-cloning and no-deleting together (and
excluding wavefunction collapse as a valid physical process) we
see that quantum information (of non-orthogonal states) has a
quality of ``permanence'': creation of copies can only be achieved
by importing the information from some other part of the world
where it had {\em already existed}; destruction (deletion of a
copy) can only be achieved by exporting the information out to
some other part of the world where it must {\em continue to
exist}. This property is different from the {\em preservation} of
information by any reversible dynamics. For example the classical
reversible C-NOT operation can imprint copies of a bit $b$ into a
standard state via $\ket{b}\ket{0} \rightarrow \ket{b}\ket{b}$ and
also delete copies via $\ket{b}\ket{b} \rightarrow \ket{b}\ket{0}$
but in both cases the first copy is used in an essential way in
the process and the information content of one copy is the same as
that of two copies. In contrast, in the quantum (non-orthogonal)
case, copying and deleting can only occur independently of the
first copy and then reversibility of dynamics implies that the
information of the second copy must have already separately
existed in the environment (for cloning) or continue to exist
separately in the environment (for deletion). But in any
reasonable intuitive sense, $\ket{\psi_i}\ket{\psi_i}$ does not
have double the information content of $\ket{\psi_i}$ and one
might interpret this as an overlap of information content of the
two copies. Then curiously, this common part cannot be taken from
a single copy and merely extended, to give the second copy.

\bigskip

\noindent {\Large\bf Acknowledgements}

\noindent RJ is supported by the U.K. Engineering and Physical
Sciences Research Council.

\end{document}